\documentstyle[preprint,prabib,aps,12pt]{revtex}
\begin{document}

\title{Matching Condition on the Event Horizon and Holography 
Principle}
\author{V. Dzhunushaliev
\thanks{E-Mail Address : dzhun@rz.uni-potsdam.de and 
dzhun@freenet.bishkek.su}}
\address{Universit\"at Potsdam, Institute f\"ur Mathematik,
14469, Potsdam, Germany \\
and Theor. Phys. Dept. KSNU, 720024, Bishkek, Kyrgyzstan}

\maketitle

\begin{abstract}
It is shown that the event horizon of 4D black hole or $ds^2 = 0$ 
surfaces of multidimensional wormhole-like solutions reduce 
the amount of information necessary for determining the whole spacetime 
and hence satisfy the Holography principle. This leads to the fact 
that by matching two metrics on a $ds^2 = 0$ surface (an event 
horizon for 4D black holes) we can match only the metric components 
but not their derivatives. For example, this allows us to obtain 
a composite wormhole inserting a 5D wormhole-like flux tube 
between two Reissner-Nordstr\"om black holes and matching them on the 
event horizon. Using the Holography principle, the entropy of a black hole 
from the algorithm theory is obtained. 
\end{abstract}
\pacs{}

\section{Introduction}
Matching two metrics which are solutions of 
the Einstein equation leads to the fact that 
surface stress-energy appears on the 
matching surface. 
This is a consequence of the Einstein equations. 
A detail explanation of this can be found, for example, in 
Refs \cite{israel66}, \cite{choquet68}, \cite{ellis95}. 
A cause of this is evident: the 
Riemann tensor contains second derivatives of the metric 
which lead to a $\delta$-function in the left-hand side of the 
Einstein equations, hence in the right-hand side there should be 
$\delta$-like surface stress-energy. 
\par 
But the Holography principle proposed in Refs 
\cite{hooft1}, \cite{thorn1}, \cite{susskind95} tells us that 
there is a surface which 
essentially cuts down the number of degrees of freedom. 
It can suggest that matching two metrics on this surface can 
substantially change the matching procedure on such a 
surface. For this purpose we propose the Lorentz invariant 
surface on which $ds^2 = 0$  
\footnote{often this is an event horizon (EH) but 
for some wormhole-like non-asymptotically flat 
multidimensional solutions  it is not so.}. 
\par 
Further we consider several solutions 
of 4D and vacuum multidimensional (MD) gravity: two solutions 
are the Reissner-Nordstr\"om and Yang-Mills black holes 
(BH) and two solutions are wormhole-like (WH) solutions 
which in some sense are dual to the above-mentioned BHs 
\footnote{This duality means that static regions of the 
4D BHs are given by $r \ge r_g$ but the static regions 
of the MD wormhole-like solutions by 
$|r| \le r_g$.}. 
\par 
To begin with, we bring the definition of the Holography 
principle according to \cite{susskind95}: 
''... a full description of nature requires only a two-dimensional 
lattice at the spatial boundaries of the 
world ...``. Our aim is to show that this principle works on the 
$ds^2 = 0$ surface.

\section{4D case}

\subsection{Event horizon as a Holography surface for 
the Reissner-Nordstr\"om BH}

The metric in this case is 
\begin{equation} 
ds^2 = \Delta (r) dt^2 - \frac{dr^2}{\Delta (r)} - 
r^2 \left (d\theta ^2 + \sin ^2 \theta d\varphi ^2 
\right ) , 
\label{2-1-1} 
\end{equation} 
the electromagnetic potential is 
\begin{equation} 
A_\mu = u = u = \{\omega (r), 0, 0, 0\} .
\label{2-1-2} 
\end{equation} 
The Einstein - Maxwell equations can be written as 
\begin{eqnarray} 
-\frac{\Delta '}{r} + \frac{1 - \Delta}{r^2} & = &
\frac{\kappa}{2} {\omega '} ^2  ,
\label{2-1-3} \\
-\frac{\Delta ''}{2} - \frac{\Delta '}{r} & = & 
-\frac{\kappa}{2} {\omega '} ^2  ,
\label{2-1-4} \\
\omega ' & = & \frac{q}{r^2}     .
\label{2-1-5}
\end{eqnarray} 
It is easy to prove that Eq. (\ref{2-1-4}) is a 
consequence of (\ref{2-1-3}) and (\ref{2-1-5}). 
For the Reissner - Nordstr\"om 
BH the event horizon (EH) is defined by the condition 
$\Delta (r_g) = 0$, where $r_g$ is the radius of the EH. 
Hence in this case we see that on the EH 
\begin{equation} 
\Delta '_g = \frac{1}{r_g} - \frac{\kappa}{2}
r_g {\omega '_g}^2  ,
\label{2-1-6} 
\end{equation}
here (g) means that the corresponding value is taken 
on the EH. Thus, Eq. (\ref{2-1-3}), which is the Einstein equation, 
is a first-order differential equations 
in the whole spacetime $(r \ge r_g)$. The condition (\ref{2-1-6}) 
tells us that the derivative of the metric on the EH is expressed 
through the metric value on the EH. And this means that 
Holography principle works here and is connected with 
the presence of the EH. In passing we remark that this 
allows us to calculate the BH entropy from an algorithmical 
point of view \cite{dzh4} without any quantum-mechanical 
calculations, which we will do in \ref{algor}.

\subsection{Event horizon as a Holography surface for an 
SU(2) Yang-Mills BH} 

\label{su2ym}

Here we use the following metric 
\begin{equation}
ds^2 = e^{2\nu(r)} \Delta (r) dt^2 - \frac{dr^2}{\Delta (r)} - 
r^2 \left (d\theta ^2 + \sin ^2 \theta d\varphi ^2 
\right ) .
\label{2-2-1} 
\end{equation}
For the SU(2) Yang-Mills gauge field we choose the following 
monopole-like ansatz as in \cite{gal1} 
\begin{eqnarray}
A^a_i & = & \frac{\epsilon ^a_{ij} x^j}{r^2} \left (1 - f(r)\right ) ,
\label{2-2-2}\\
A^a_t & = & \frac{x^a}{r^2} v(r)  ,
\label{2-2-3}
\end{eqnarray}
here $a=1,2,3$ is the internal index; $i = 1,2,3$ is the spatial 
index. For simplicity we consider the case $v=0$. Thus, we have 
the following set of Einstein - Yang - Mills equations 
\begin{eqnarray}
-\frac{\Delta '}{r} + \frac{1 - \Delta}{r^2} & = &
\frac{\kappa}{r^2} \left [\Delta {f'}^2 + \frac{1}{2 r^2} 
\left (f^2 - 1\right )^2\right ]  ,
\label{2-2-4}\\
\nu ' & = & \frac{\kappa}{r} {f'}^2   ,
\label{2-2-5}\\
R^\theta_\theta - \frac{1}{2} R & = & \kappa T^\theta_\theta  ,
\label{2-2-6}\\
\Delta f'' + \Delta f'\nu ' + f'\Delta ' & = & \frac{f}{r^2} 
\left (f^2 - 1\right ) .
\label{2-2-7}
\end{eqnarray}
Due to the Bianchi identities Eq. (\ref{2-2-6}) is a 
consequence of the other equations. From 
(\ref{2-2-5}) we have
\begin{equation}
\nu (r)= \kappa \int_{r}^{\infty} \frac{{f'}^2}{r}dr
\label{2-2-8} ,
\end{equation}
here we choose the time so that $\nu_{r\to \infty} = 0$. 
So the value on the EH is 
\begin{equation}
\nu_g = \kappa \int_{r_g}^{\infty} \frac{{f'}^2}{r}dr .
\label{2-2-9}
\end{equation}
The presence of an EH means that $\Delta (r_g) = 0$, hence 
close to an EH 
\begin{eqnarray}
\Delta & = & \Delta _1 x + \Delta _2 \frac{x^2}{2} + \cdots   ,
\label{2-2-10}\\
f & = & f_0 + f_1x + f_2\frac{x^2}{2} + \cdots  ,
\label{2-2-11}
\end{eqnarray}
here $x = r-r_g$. Then from the Einstein - Yang - Mills equations 
we have: 
\begin{eqnarray}
f_2 & = & -\frac{\kappa}{r_g} f_1^3  ,
\label{2-2-12}\\
f_1 & = & \frac{f_0}{r_g^2 \Delta _1} \left( f_0^2 - 1 \right)  ,
\label{2-2-13}\\
\Delta _1 & = & \frac{1}{r_g} - \frac{\kappa}{2 r_g^3} 
\left( f_0^2 - 1 \right)^2 .
\label{2-2-14}
\end{eqnarray}
Thus we have only one physically significant parameter 
$f_0 = f(r_g)$\footnote{we can exclude $r_g$ making all 
the magnitudes dimensionless 
$(r^* = r/r_g$; $\kappa ' = \kappa /r_g^2$).}. Again we have the Holography 
principle on the EH. In this case even more: the Yang - Mills equation 
satisfy the Holography principle since the first derivative 
of $f(r)$ on the EH 
is expressed through the value of $f_0$ on the EH, which is seen from 
Eq. (\ref{2-2-13}). 

\section{Multidimensional wormhole-like cases}

\subsection{$ds^2 = 0$ surface as Hologram surface for 
5D WH-like solution}

Let us consider the 5D WH-like metric 
\begin{equation}
ds^2 = \frac{1}{\Delta (r)} dt^2 - R_0^2 \Delta (r) 
\biggl [ d \chi + \omega (r) dt \biggl ]^2 - dr^2 - 
a(r)\left( d\theta ^2 + \sin ^2\theta d\varphi ^2 \right) ,
\label{3-1-1}
\end{equation}
here $\chi$ is the 5$^{th}$ coordinate, $r$, $\theta$ and 
$\varphi$ are the ordinary spherical coordinates,  
$R_0$ is some constant. The 5D Einstein equations are 
\begin{eqnarray}
\Delta \Delta '' - {\Delta '}^2 + \Delta \Delta ' \frac{a'}{a} 
+ R_0^2 \Delta ^4{\omega '}^2 & = & 0  ,
\label{3-1-2}\\
\left( a\Delta ^2 \omega ' \right)' & = & 0  ,
\label{3-1-3}\\
a'' & = & 2  .
\label{3-1-4}
\end{eqnarray}
The solution is \cite{dzh16} 
\begin{eqnarray}
a & = & r^2 + r^2_0
\label{3-1-5}\\
\Delta & = & {q \over 2r_{0}}{r^{2}_{0} - r^{2}
\over r^{2}_{0} + r^{2}},
\label{3-1-6}\\
\omega & = & 4r^{2}_{0}\over q}
{r\over {r^{2}_{0} - r^{2}} .
\label{3-1-7}
\end{eqnarray}
here $q$ and $r_0$ are some constants. 
It is easy to prove that 
$G_{tt}(\pm r_0) = \Delta ^{-1}(\pm r_0) - R_0^2 \Delta (\pm r_0)
\omega ^2(\pm r_0) = 0$ 
and $ds^2 = 0$ on surfaces $r = \pm r_0$. In this sense the surface 
$r = r_0$ is analogous to an EH. On the surface $ds^2 = 0$ 
$\Delta (\pm r_0) = 0$, and therefore from Eq. 
(\ref{3-1-2}) we have 
\begin{equation}
\Delta '_0 = \pm \frac{q}{a_0} = \pm \frac{q}{2r_0^2}
\label{3-1-8} .
\end{equation}
The signs $(\pm)$ correspond, respectively, to $(r = \mp r_0)$ where the 
surfaces $ds^2 = 0$ are located. This also indicates that the 
surface $ds^2=0$ works here according to the Holography principle.

\subsection{The surface $ds^2 = 0$ as a Holography surface for a 
7D WH-like solution}

Here we work with gravity on the principal bundle 
as in Ref. \cite{dzh2}, i.e. the base of the bundle is an 
ordinary 4D Einstein spacetime and the fibre of the bundle 
is the SU(2) gauge group. In our case we have gravity on the 
SU(2) principal bundle with the SU(2) structural group (simultaneously 
it gives the extra coordinates). This group as the sphere $S^3$ is the 
space of the extra dimensions. Thus, the dimension of our MD gravity 
is 7.
\par 
The gravity equations are: 
\begin{eqnarray}
R_{a\mu} & = & 0    ,
\label{3-2-2}\\
R^a_a =  R^4_4 + R^5_5 + R^6_6 & = & 0 ,
\label{3-2-3}
\end{eqnarray}
here $A = 0,1,2, \ldots ,6$ is a MD index on the total space of the 
bundle, $\mu = 0,1,2,3$ is the index on the base of the bundle, 
$a = 4,5,6$ is the index on the fibre of the bundle. 
For MD gravity on the principal bundle we have the following 
theorem \cite{per1,sal1}: 
\par
Let $G$ be the group 
fibre of the principal  bundle.  Then  there  is a one-to-one
correspondence between the $G$-invariant metrics on the  total  
space ${\cal X}$
and the triples $(g_{\mu \nu }, A^{a}_{\mu }, h\gamma _{ab})$. 
Here $g_{\mu \nu }$ is Einstein's pseudo  -
Riemannian metric on the base; $A^{a}_{\mu }$ is the gauge field 
of the group $G$ ( the nondiagonal components of 
the multidimensional metric); $h\gamma _{ab}$  is the 
symmetric metric on the fibre. 
\par 
According to this theorem 7D metric has the following form 
\begin{equation}
ds^2 = \frac{\Sigma ^2(r)}{u^3(r)}dt^2 - R_0^2 u(r) 
\left( \sigma ^a + A^a_\mu dx ^\mu \right)^2 - d r^2 - 
a(r)\left( d \theta ^2 + \sin\theta d\varphi ^2 \right) ,
\label{3-2-1}
\end{equation}
here $A^{a}_\mu$ is the above-mentioned SU(2) gauge field, 
the one-forms $\sigma ^{a}$ on the SU(2) group can be written as follows:
\begin{eqnarray}
\sigma ^{1} & = & {1\over 2}
(\sin \alpha d\beta - \sin \beta \cos \alpha d\gamma ),
\label{3-4-1}\\
\sigma ^{2} & = & -{1\over 2}(\cos \alpha d\beta +
\sin \beta \sin \alpha d\gamma ),
\label{3-4-2}\\
\sigma ^{3} & = & {1\over 2}(d\alpha +\cos \beta d\gamma ),
\label{3-4-3}
\end{eqnarray}
here we have introduced Euler's angles 
$\alpha ,\beta , \gamma$ on the fibre ($SU(2)$ group)  
and $0\le \beta \le \pi , 0\le \gamma \le 2\pi , 0\le \alpha \le 4\pi $. 
\par 
An ansatz for the 
gauge potential $A^a_\mu$ is taken as for the monopole 
(as in  section \ref{su2ym}). 
For simplicity we examine the case $f(r)=0$ only. Then we have 
the following vacuum gravity equations 
\begin{eqnarray}
v'' - \frac{\Sigma 'v'}{\Sigma} + 4 \frac{u'v'}{u} +
\frac{a'v'}{a} & = & 0  ,
\label{3-2-4}\\
\frac{a''}{a} + \frac{a'\Sigma '}{a\Sigma} - \frac{2}{a} + 
\frac{R_0^2 u}{4a^2} & = & 0  ,
\label{3-2-5}\\
\frac{u''}{u} + \frac{\Sigma 'u'}{\Sigma u} - \frac{{u'}^2}{u^2} + 
\frac{a'u'}{au} - \frac{4}{R^2_0u} - \frac{1}{12}\frac{R^2_0u}{a^2} + 
\frac{1}{3}\frac{R^2_0u^4}{\Sigma ^2}{v'}^2 & = & 0  ,
\label{3-2-6}\\
\frac{\Sigma ''}{\Sigma} + \frac{a'\Sigma '}{a\Sigma} - 
\frac{6}{R^2_0u} - 
\frac{R^2_0u}{8a^2} & = &0  .
\label{3-2-7}
\end{eqnarray}
In Ref. \cite{dzh1} an approximative solution has been found. Here 
we are interested only in what happens close to the surface 
$ds^2(\pm r_0) = 0$. In order that a surface $ds^2(\pm r_0) = 0$ 
exist it is necessary that the following condition be satisfied: 
\begin{equation}
G_{tt} = \frac{\Sigma ^2_0}{u^3_0} - R_0^2 u_0 v_0^2 = 0 ,
\label{3-2-10}
\end{equation}
here the index (0) means that the corresponding quantities are taken
on the surface $r=\pm r_0$. 
We suppose that in this region there is the following behaviour 
($r = +r_0$) 
\begin{eqnarray}
u(r) & = & u_0 \left( 1 - \frac{r}{r_0} \right)^{1/2} + \cdots   ,
\label{3-2-8}\\
\Sigma (r)& = & \Sigma _0 + \Sigma _1\left( 1 -  \frac{r}{r_0} 
\right)^{3/2} 
+ \cdots  ,
\label{3-2-9} \\
v(r) & = & \frac{q\Sigma _0}{a_0 u^4_0}\frac{1}{1 - \frac{r}{r_0}} + 
\cdots  .
\label{3-2-9a}
\end{eqnarray}
This leads to the following result: 
\begin{eqnarray}
u_0^2 & = & \sqrt{\frac{2}{3}} \frac{|q|r_0}{a_0}  , 
\label{3-2-11}\\
R_0 & = & \sqrt {\frac{2}{3}} u_0^2 r_0  ,
\label{3-2-12}\\
\frac{\Sigma _1}{\Sigma _0} & = & \frac{12}{u^5_0}  .
\label{3-2-13}
\end{eqnarray}
Here we also see the Holography principle: $u_0$ and 
$\Sigma ' (r_0) = \Sigma _1$ are 
not undependent initial data, they are determinated from the dimensionless 
magnitudes $q/r_0$ and $a_0/r^2_0$. 

\section{Discussion}

Thus we see that at least for static spherically symmetric 
solutions 
in 4D and vacuum MD gravity the Holography principle leads from 
the presence of the $ds^2 = 0$ surface (event horizon for the 4D 
gravity). 
For researchers working with 4D Einstein - Yang - Mills black holes 
this is well known: the condition (\ref{2-2-13}) is necessary for 
numerical calculations (see, for example, \cite{gal1}). 
\par 
These results allow us to say that (at least for the static 
spherically symmetric cases) on the surface $ds^2 = 0$ the Holography 
principle changes and simplifies the matching conditions 
due to reduction of the physical degrees of freedom.  
Roughly speaking, close to this surface the Einstein differential 
equations of 
the second  order are reduced to first-order equations\footnote{for 
the 4D spherically symmetric case it is an exact result.}. 
In this case it is evident that: 
\textit{matching of two metrics on the surface $ds^2 = 0$ 
does not lead to $\delta-$functions and hence the appearance 
of an additional surface stress-energy.} 

\subsection{Composite WH with 5D WH-like solution 
and two Reissner-Nordstr\"om black holes}

For example, having a 5D WH-like solution, we can match to it two 
Reissner-Nordstr\"om black holes on the two $ds^2 = 0$ 
surfaces \cite{dzh7}. This can be done since ordinary 4D electrogravity can 
be 
considered as 5D vacuum gravity in the initial Kaluza 
sense\footnote{when $G_{55} = 1$ and the Lagrangian is not varied with 
respect to $G_{55}$.} and on the EH we join fibre to fibre and base to 
base 
Reissner - Nordstr\"om and 5D WH-like solutions. 
In this case we have to match on the surface $ds^2 = 0$ 
(an EH for an observer at infinity) only the following 
quantities  
\begin{itemize}
 \item 
 The area of the surface $ds^2 = 0$ of the 5D WH-like solution with the 
area of the EH 
 of the Reissner - Nordstr\"om BH: 
 \begin{equation}
 4\pi a = 4\pi r^2_g ,
 \label{4-1-1}
 \end{equation}
 here the left-hand side is 5D and the right-hand side is 4D. 
 \item 
 Let we compare the 5D equation $R_{15} = 0$ 
 \begin{equation}
 \left( 4\pi a \omega ' e^{-2\nu} \right)' = 0 ,
 \label{4-1-2}
 \end{equation}
 with the Maxwell equation 
 \begin{equation}
 \left( 4\pi r^2 E \right)' = 0 .
 \label{4-1-3}
 \end{equation}
 In both cases $4\pi a$ or $4\pi r^2$ is the area of an 2-sphere and 
 Eqs (\ref{4-1-2}) and (\ref{4-1-3}) tell us that the electrical 
 field flux is preserved. Hence we can make a conclusion that 
 $\omega ' e^{-2\nu}$ is a ''5D electrical`` field, 
 $E_5 = \omega ' e^{-2\nu}$ and for the 4D case we have the conventional 
 definition of the electrical field $E$. Hence on the $ds^2 = 0$ matching 
surface 
 \begin{eqnarray}
 \omega _0' e^{-2\nu _0} = E_g ;
 \label{4-1-4}
 \end{eqnarray} 
 here (0) means that the corresponding 5D quantity is on the $ds^2 = 0$ 
 surface and (g) means that this 4D quantity is taken on the EH. 
  \item 
  We do not match $G_{rr}$ and $g_{rr}$ since these components of 4D 
and 5D 
  metrics are arbitrary: they depend only on the choice of the radial 
  coordinate. 
\end{itemize} 
In fact we see that we have only two matching conditions and this 
is evident: a Reissner - Nordstr\"om BH is characterized only by two 
physical 
quantities: the electrical charge $Q$ and the mass $m$. 
Also the 5D WH-like solution (\ref{3-1-5}) - (\ref{3-1-7}) 
is characterized only by two physical quantities: the constants
$q$ and $r_0$. It is remarkable that for these 4D and 5D physical quantities 
we have only the matching conditions (\ref{4-1-1}) and (\ref{4-1-4}) 
as a consequence of the Holography principle. 

\subsection{The Holography principle and algorithmical complexity}
\label{algor}

It is interesting that reduction of the order of Einstein's 
differential equations near an EH and the Holography 
principle allows us to calculate the entropy of a BH without 
any quantum calculations. We shortly repeat this result 
obtained in \cite{dzh4}. 
In the 1960's Kolmogorov ascertained that the algorithm theory 
allows us to define the probability notion for a single object. 
His idea is very simple: the probability is connected with 
the complexity of this object, ''chance`` = ''complexity``. 
The more complex (longer) is an algorithm describing this 
object\footnote{such an algorithm can, for example, consist in 
the field equations 
describing a field distribution in the spacetime.} 
the smaller probability it has. Of course there is the question: 
what is it the length of an algorithm ? It was found that 
such an invariant, well posed definition can be given \cite{kol1}: 
\par
The algorithmic complexity ${\cal K}(x\mid y)$ of the  object $x$  for
a given object $y$ is the minimal length of the "program" $P$
which is written as a sequence of  the  zeros  and  unities
and allows one to construct $x$ from given $y$
\par
\begin{equation}
{\cal K} (x\mid y) = \min_{A( P,y)=x} l(P)
\label{4-2-1}
\end{equation}
where $l(P)$ is the length of the  program $P$; $A(P,y)$  is  the 
algorithm calculating the object $x$, using  the  program $P$ when 
the object $y$ is given. 
Then we can determine the algorithmical complexity of a BH 
and the logarithm of it gives us the entropy of BH. 
\par 
We write the initial equations for describing the Schwarzschild BH. The 
metric is 
\begin{equation}
ds^{2} = dt^{2} - e^{\lambda (t,R)}dR^{2} - r^{2}(t,R) \left( d\theta 
^{2} +
\sin ^{2} \theta d\phi ^{2} \right) ,
\label{4-2-2}
\end{equation}
here $t$ is time, $R$ is radius, $\theta$ and $\phi$ are  polar  angles. 
In this case Einstein's equations are
\begin{eqnarray}
-e^{-\lambda}r'^2 + 2r\ddot r + \dot r ^2 + 1 & = & 0,
\label{4-2-8}\\
-\frac{e^{-\lambda}}{r}\left (2r'' - r'\lambda ' \right ) + 
\frac{\dot r \dot \lambda}{t} + \ddot \lambda + 
\frac{\dot \lambda ^2}{2} + 
\frac{2\ddot r}{r} & = & 0,
\label{4-2-9}\\
-\frac{e{-\lambda}}{r^{2}}
 \left(2rr'' + {r'} ^{2} -rr'\lambda '\right) +
 \frac{1}{r^{2}}\left( r\dot a \dot \lambda + {\dot a} ^{2} + 
1 \right) & = & 0,
\label{4-2-10}\\
2\dot r' - \dot \lambda r' & = & 0,
\label{4-2-11}
\end{eqnarray}
where $(')$ and $(\dot{\phantom x})$ mean, respectively, derivatives
in $t$ and $r$. The $0\choose 0$ Einstein's equation for the initial 
data is 
\begin{equation}
-\frac{e^{-\lambda}}{r^{2}}
 \left(2rr'' + {r'} ^{2} -rr'\lambda '\right) +
 \frac{1}{r^{2}}\left( r\dot a \dot \lambda + 
{\dot a} ^{2} + 1 \right) = 0 .
\label{4-2-3}
\end{equation}
The Caushy  hypersurface determining the whole Schwarzschild - 
Kruskal spacetime is $t = 0$. The ''quantity`` of the initial data 
according to the Holography principle can be essentially reduced:  
first, at $t = 0$ the first time derivative of all metric components 
is equal to zero. Therefore we have the following equation for the 
initial data  
\begin{equation}
2rr''  + {r'} ^{2} - rr' \lambda '  - e^{\lambda } = 0 .
\label{4-2-4}
\end{equation}
We know that the hypersurface $t = 0$  
is a WH connecting two asymptotically flat, causally disconnected 
regions. This WH is symmetrical relative to centre $r = r_g$, 
therefore the initial data for this equation are 
\begin{eqnarray}
r' (R=0,t=0) = 0 ,
\label{4-2-5}\\
r(R=0,t=0) = r_{g} ,
\label{4-2-6}
\end{eqnarray}
where $r_{g}$ is the radius of the event horizon. The conditions 
(\ref{4-2-5}) and (\ref{4-2-6}) 
are necessary for this WH to exist and 
this is also a consequence of Holography principle (reducing 
the information describing the BH). Thus, for describing the whole 
Schwarzschild - Kruskal spacetime we need the algorithm 
(\ref{4-2-8}) - (\ref{4-2-11}) and the initial data (\ref{4-2-6}). 
Therefore the algorithmical complexity of the Schwarzschild 
BH $\cal K$ is defined by the given expression 
\begin{equation}
{\cal K} \approx  L_{initial}\left( \frac{r_{g}}{r_{Pl}} \right) ^2 + 
L_{Einstein \; equations}  ,
\label{4-2-7}
\end{equation}
where $L_{initial}$ is  the length of the algorithm 
(program) determining the dimensionless number 
$r^2_{g}/r_{Pl}^2$, performed on some universal 
machine, $L_{Einstein \; equations}$ is the length of the 
algorithm (program) for solving the Einstein equations 
(\ref{4-2-8}) - (\ref{4-2-11}) using some universal machine, 
for example, the Turing machine. 

\section{Conclusion}

Finally, we can suppose that the event horizon plays an exceptional 
role in nature: \textit{it is the surface on which the Holography 
principle is realized}. We have shown that in this case  
an EH can \textit{divide} the regions in our Universe 
with splitting off and nonsplitting off the extra dimensions in the above - 
mentioned sense as a consequence of realizing the Holography principle.

\section{Acknowledgments}

This work is supported by a Georg Forster Research Fellowship
from the Alexander von Humboldt Foundation. I would like to
thank H.-J. Schmidt for the invitation to
Potsdam Universit\"at for research, D.Singleton 
and K. Bronnikov for a fruitful discussion.


\end{document}